\documentclass[%
reprint,
 amsmath,amssymb,
 aps,
]{revtex4-1}

\usepackage{graphicx}
\usepackage{dcolumn}
\usepackage{bm}

\usepackage{soul}
\usepackage{cancel}
\usepackage{xspace}
\usepackage{lineno}
\usepackage{lineno,hyperref}
\usepackage{xcolor}

\newcommand{\pt}{\ensuremath{p_{\rm T}}\xspace}

\newcommand{\Qppb}{\ensuremath{Q_{\rm pPb}}\xspace}

\newcommand{\py}{PYTHIA\xspace}
\newcommand{\pya}{PYTHIA/Angantyr\xspace}
\newcommand{\pPb}{p--Pb\xspace}
\newcommand{\pp}{pp\xspace}

\begin{document}

\preprint{APS/123-QED}
\title{Flattenicity as ``centrality'' estimator in p-Pb collisions \\ simulated with PYTHIA 8.312 Angantyr}%

\author{Antonio Ortiz}
\email{antonio.ortiz@nucleares.unam.mx}
\affiliation{%
Instituto de Ciencias Nucleares, UNAM,
 Apartado Postal 70-543, Ciudad de M\'exico 04510, M\'exico 
}

\author{Gyula Benc\'edi}
\affiliation{HUN-REN Wigner Research Centre for Physics, Budapest, Hungary}

\author{Feng Fan}
\affiliation{Central China Normal University, Wuhan, Hubei, 430079, China}

\date{\today}

\begin{abstract}
In this paper, a ``centrality'' estimator based on  flattenicity ($\rho$) is studied in proton-led (p-Pb) collisions at $\sqrt{s_{\rm NN}}=5.02$\,TeV using PYTHIA~8.312 Angantyr. Although Angantyr is still under development, the existing implementation is enough to study the particle production in dense QCD systems where medium effects are absent. Firstly, ALICE data on pseudorapidity distributions as a function of the forward multiplicity (V0M), as well as transverse momentum distributions of identified particles in non-single diffractive p--Pb collisions, are compared with the most recent version of Angantyr. Secondly, the average number of binary nucleon-nucleon ($N_{\rm coll}$) collisions for different ``centrality'' estimators are compared. The studies include the following ``centrality'' estimators: V0M, $\rho$ and midrapidity multiplicity (CL1). On one hand, the ``centrality'' dependence of $\langle N_{\rm coll} \rangle$ for the $\rho$ selection shows the smallest deviations ($<8$\%) with respect to that obtained using impact parameter $b$; on the other hand, the V0M and CL1 yield huge deviations (up to a factor 2) with respect to the results using $b$. The particle ratios (proton-to-pion and kaon-to-pion ratios) and nuclear modification factors ($Q_{\rm pPb}$) as a function of $p_{\rm T}$ are also studied for the different ``centrality'' estimators. The proton-to-pion ratio exhibits a flow-like peak at intermediate $p_{\rm T}$ (2-8\,GeV/$c$) with little or no ``centrality'' dependence for V0M, $\rho$ and $b$ selections. The kaon-to-pion ratio as a function of $p_{\rm T}$ is ``centrality'' independent for the same selections. On the contrary, for the high-multiplicity CL1 class the ratios exhibit the typical behaviour associated with hard physics. Regarding $Q_{\rm pPb}$, a peak at intermediate $p_{\rm T}$ ($2-8$\,GeV/$c$) for different particle species is observed when the ``centrality'' is obtained with $b$ or $\rho$. The observed features diminish for the selections based on V0M and CL1. The studies presented here are relevant to help in the investigation of the plethora of effects, which have been reported in small systems by experiments at the LHC.

\end{abstract}

\maketitle


\section{\label{sec:1}Introduction}

One of the main results of the last decade is the discovery of heavy-ion-like behaviour in small-collision systems  (\pp and \pPb collisions) at the Large Hadron Collider (LHC)~\cite{Nagle:2018nvi}. The phenomena include strangeness enhancement~\cite{ALICE:2016fzo} and collective-like behaviour~\cite{CMS:2016fnw}. The origin of the new effects remains as an open question. In particular, because in heavy-ion collisions there is strong evidence that supports the formation of the strongly-interacting quark--gluon plasma (sQGP) in central collisions, and there, jet quenching effects have been observed but they are absent in small-collision systems~\cite{Busza:2018rrf,ALICE:2022wpn}. 

Based on  IP-Glasma calculations the energy density,  averaged over the transverse area, in high-multiplicity pp collisions at the LHC energies can be around 70\,GeV/fm$^{3}$ for charged-particle multiplicity densities above 100~\cite{ALICE:2020fuk}. From lattice quantum chromodynamics (QCD) calculations, this condition would be enough to expect the formation of the strongly-interacting quark--gluon plasma (QGP) in pp collisions. However, no signatures of jet quenching have been observed so far~\cite{ALICE:2022qxg}. Moreover, long-range angular correlations have been observed even in very dilute systems like low-multiplicity pp collisions~\cite{ALICE:2023ulm}, photonuclear ultra-peripheral Pb--Pb collisions~\cite{ATLAS:2021jhn}, and probably within jets~\cite{CMS:2023lgq}. Although hydrodynamics has been applied to \pp and \pPb collisions and it describes some aspects of data, theoretical works suggest that it should not be applicable in small-collisions systems~\cite{Werthmann:2023dvl}. 
This opens the door for alternative approaches like colour strings that have successfully described some aspects of pp data~\cite{Bierlich:2014xba}. A recent implementation of heavy-ion collisions in \py (Angantyr) opened a new venue to explore the dense QCD systems without QGP effects. In particular, it is amazing that with a simplistic extrapolation of the pp dynamics to heavy-ion collisions, \py (Angantyr) is even able to describe the low transverse momentum yield ($p_{\rm T}<3$\,GeV/$c$) in central heavy-ion collisions within less than 20\%~\cite{Bierlich:2018xfw}. The model is still far to be completed, therefore, any physics conclusion has to be taken with a grain of salt. However, given that \py does a relatively good job in describing general aspects of LHC data, and the affordable computing resources needed to run simulations, the present paper uses this Monte Carlo event generator to explore new ways to analyse p--Pb collisions. 

In heavy-ion collisions, the observables are typically studied as a function of the collision centrality. Centrality is related with the impact parameter $b$ of the collision. For heavy-ion collisions, the centrality is well defined given the strong correlation between the impact parameter of the collision and the measured event activity (particle multiplicity). However, this is not the case for diluted systems like those created in pp and p--Pb collisions. Because there, the contribution of the jet fragments to the particle multiplicity becomes relevant. This issue has been recently addressed using ALICE data~\cite{ALICE:2023plt}, there it has been shown that even requiring high multiplicities at forward-backward pseudorapidities selects pp collisions with an energetic jet in the forward detector, consequently biasing the particle production measured in the central region.

The Glauber model~\cite{Miller:2007ri} is generally used to calculate geometrical quantities of nuclear collisions (A--A or p--A). In this model, the impact parameter controls the average number of participating nucleons, $N_{\rm part}$ and the corresponding number of binary collisions $N_{\rm coll}$. It is expected that variations of the amount of matter overlapping in the collision region will change the number of produced particles, and parameters such as $N_{\rm part}$ and $N_{\rm coll}$ have traditionally been used to describe those changes quantitatively, and to relate them to minimum-bias pp collisions.

Typically, the centrality estimators for A--A collisions are based either on multiplicity or summed energy in the pseudorapidity intervals defined by detectors. However, as pointed out this Ref.~\cite{ALICE:2014xsp}, applying the same approach to p--Pb collisions is not straightforward, namely because selecting high (low) multiplicity p--A collisions biases the sample towards  large (small) average $N_{\rm part}$, and also positive (negative) multiplicity fluctuations leading to deviations from the binary scaling of hard processes. These fluctuations are partly related to qualitatively different types of collisions. For example, high multiplicity pp (or p--Pb) collisions show a continuous increase of the high-\pt particle yield relative to minimum-bias pp collisions when both \pt spectra and multiplicity are measured at midrapidity~\cite{ALICE:2019dfi}. In order to reduce the biases in p--Pb collisions, the ALICE collaboration proposed to determine the collision centrality measuring the forward-neutron energy~\cite{ALICE:2014xsp}. This approach works because given the large rapidity gap, the auto correlations are expected to be significantly reduced.  In this paper, a new ``centrality'' estimator based on flattenicity is proposed to study the p--Pb collisions, flattenicity was originally proposed to study pp collisions~\cite{Ortiz:2022zqr,Ortiz:2022mfv}, and the first ALICE data are already available~\cite{Bencedi:2023iib}. The results are compared with those obtained using other ``centrality'' estimators based on particle multiplicity both at midrapidity and at forward pseudorapidity.  

The paper is organised as follows. The ``centrality'' estimators are introduced in section~\ref{sec:2}. Section~\ref{sec:3} describes the main features of \pya, and comparisons between the model and ALICE data. Section~\ref{sec:4} presents the results and discussion. Finally, results with an outlook for the upcoming LHC Run 3 is presented in section~\ref{sec:5}.

\section{\label{sec:2}``Centrality'' estimators}

The event classifiers listed below are calculated for p--Pb collisions at $\sqrt{s_{\rm NN}}=5.02$\,TeV simulated with PYTHIA~8.312~Angantyr (\pya from now on). Primary-charged particles~\cite{ALICE_PUBLIC_2017_005} were considered in the calculation of the event activity estimators. The primary-charged particles are defined as particles with a mean proper lifetime larger than 1\,cm$/c$, which are either produced directly in the interaction or from decays of particles with mean proper lifetime smaller than 1\,cm$/c$. All primary-charged particles with $\pt > 0$ are considered in the calculation, the pseudorapidity cuts depend on the corresponding experimental definition. It is worth mentioning that for p--Pb collisions, the nucleon-nucleon centre-of-mass system is moving in the laboratory frame with a rapidity of $y_{\rm NN}= -0.465$ in the direction of the proton beam rapidity. In the following, $y_{\rm lab}$ ($\eta_{\rm lab}$) are used to indicate
the (pseudo)rapidity in the laboratory reference frame, whereas $y$ ($\eta$) denotes the (pseudo)rapidity in the centre-of-mass reference system where the Pb beam is assigned a positive-sign rapidity.

{\bf Midrapidity multiplicity (CL1):} this is the most basic event classifier that is defined as the number of primary-charged particles in the pseudorapidity interval of $-0.3<\eta<0.3$. Since the largest multiplicity reach is only attainable using the midrapidity multiplicity estimator, for this event classifier the largest energy density is achieved~\cite{ALICE:2020fuk}. However, for this event classifier the largest bias towards hard pp collisions is expected.

{\bf Forward multiplicity (V0M):} this event classifier considers the particle multiplicity registered in the forward pseudorapidity region that is experimentally accessible to the ALICE detector. Only primary-charged particles entering in any of the sectors covered by the VZERO detector are considered. Table~\ref{tab:1} displays the segmentation of the VZERO detector that matches the selection considered in the present study.
\begin{table} [!hpt]
                \centering 
                \caption{Pseudorapidity intervals covered by the different rings of the VZERO detector of ALICE.
                 \label{tab:1}}
                \begin{tabular}{|c |c | c |}
                \hline
                \textbf{Ring}  & \textbf{VZEROC} & \textbf{VZEROA} \\ 
                \hline
                1       & $-3.7<\eta_{\rm lab}<-3.2$  & $4.5<\eta_{\rm lab}<5.1$  \\
                2       & $-3.2<\eta_{\rm lab}<-2.7$  & $3.9<\eta_{\rm lab}<4.5$  \\
                3       & $-2.7<\eta_{\rm lab}<-2.2$  &  $3.4<\eta_{\rm lab}<3.9$ \\
                4       & $-2.2<\eta_{\rm lab}<-1.7$  &  $2.8<\eta_{\rm lab}<3.4$ \\
                \hline
                \end{tabular}
\end{table}

\begin{table*} [t]
                \centering 
                \caption{The different event classifiers explored by experiments at the LHC, the pseudorapidity intervals considered for the calculation of the classifiers are specified in the table.
                 \label{tab:2}}
                \begin{tabular}{|l |c | c | c | c | c | } 
                \hline
                \hline
                \textbf{Event classifier}  &   \textbf{Symbol} & \textbf{$\mathbf{\eta}$ coverage}  & \textbf{$\varphi$ coverage (rad)} &  \textbf{is $\eta$ used ?} & \textbf{is $\varphi$ used ?} \\ 
                \hline
                 Midrapidity multiplicity  &  $N_{\rm ch}$   & $-0.3<\eta<0.3$  & $2\pi$ & yes & no \\
                 Forward multiplicity     & V0M   & $-3.7<\eta_{\rm lab}<-1.7$ and $2.8<\eta_{\rm lab}<5.1$  & $2\pi$ & yes & no \\
                 Flattenicity      &   1-$\rho$ & $-3.7<\eta_{\rm lab}<-1.7$ and $2.8<\eta_{\rm lab}<5.1$  & $2\pi$ & yes & yes \\
                \hline
                \hline
                \end{tabular}
\end{table*}
    
{\bf Flattenicity ($\rho$):} inspired by the recently introduced flattenicity~\cite{Ortiz:2022zqr}, that was originally proposed as a new observable to be measured in the next-generation heavy-ion experiment at CERN (ALICE~3) in the LHC Runs~5 and 6~\cite{ALICE:2803563,Alfaro:2024sxc}. Flattenicity was redefined~\cite{Ortiz:2022mfv} in order to classify events using the existing detectors of ALICE~\cite{ALICE:2014sbx}, the new definition goes like:

\begin{equation}
\rho=\frac{\sqrt{\sum_{i}{(N_{\rm ch}^{{\rm cell},i} - \langle N_{\rm ch}^{\rm cell} \rangle)^{2}}/{\rm N_{cell}^{2}}}}{\langle N_{\rm ch}^{\rm cell} \rangle},
\end{equation} 
where, $N_{\rm ch}^{{\rm cell},i}$ is the average multiplicity in the elementary cell $i$ and $\langle N_{\rm ch}^{\rm cell} \rangle$ is the average of $N_{\rm ch}^{{\rm cell},i}$ in the event. Flattenicity is calculated in the pseudorapidity intervals shown in table~\ref{tab:1} and using primary-charged particles with $p_{\rm T}>0$. The additional factor $1/\sqrt{\rm N_{cell}}$ guarantees flattenicity to be smaller than unity. Moreover, in order to have a similar meaning of the limits of the new event shape to those used so far (e.g. spherocity~\cite{Ortiz:2017jho}), this paper reports results as a function of $1-\rho$. Based on the multiparton interactions (MPI) model implemented in \py, pp collisions with multijet topologies (MPI) have small flattenicity values ($1 - \rho \rightarrow 1$), whereas pp collisions with a few MPI have large flattenicity values ($1 - \rho \rightarrow 0$). Therefore, the lower bound of $1 - \rho$ aims at selecting  pp collisions with small multiplicities (a few VZERO sectors are activated) or those with jet topologies in the VZERO detector. In contrast, the upper bound of $1 - \rho$ is associated with events with spherical topologies that contain particles from several multiparton interactions, which means high multiplicities. 

The characteristics of the estimators are depicted in table~\ref{tab:2}.

\section{\label{sec:3}The \pya model}

In the studies presented in this paper, the PYTHIA 8.312 Angantyr model  is used. The \pya model is an extrapolation of the pp dynamics to collisions with nuclei with a minimal set of adjustable parameters that tries to describe the general features of the final state in p--A and A--A collisions, such as multiplicity and transverse momentum distributions, without including a hydrodynamic evolution. It is based on the Fritiof model~\cite{Andersson:1986gw} and the number of participating nucleons is calculated using the Glauber formalism. The formalism also considers dynamic colour fluctuations~\cite{Alvioli:2014sba,Bierlich:2019wld}, i.e. the size of the nucleons is allowed to change in every collision, and takes into account that the colour state of the projectile must remain frozen throughout the passage of the target nucleus. These effects, in turn, will result in large fluctuations in the cross sections.

It does not assume a thermalised medium at high temperature, the motivation of using this model is to test the sensitivity of flattenicity to the impact parameter of the collision. 

In \pya the number of wounded nucleons (participant nucleons) is calculated from the Glauber model in impact parameter space, including the so-called Gribov corrections~\cite{Gribov:1968jf,Heiselberg:1991is, Blaettel:1993ah, Alvioli:2013vk} due to diffractive excitation of individual nucleons.  The primary collision is modelled as a normal inelastic, non-diffractive collision, whereas the secondary absorptive one is treated with inspiration from the Fritiof model.  In \pya the subsystems are allowed to have MPIs in secondary collisions, following the interleaved MPI shower used in \py. This procedure can be generalised to an arbitrary A--A collision. In \pya this is done by first ordering all possible interactions in increasing local impact parameter. Going from smallest to largest impact parameter, an interaction is labelled primary if neither of the participating nucleons have participated in a previous interaction and secondary otherwise.

On the conceptual level, there are some differences between pp collisions generated with the MPI-based model of \py and p--A or A--A collisions generated with \pya. Most importantly, for collisions involving heavy nuclei, colour reconnection (CR), which has been made available only recently~\cite{Lonnblad:2023stc,Bierlich:2023okq}, is only applied at the level of individual nucleon–nucleon collisions. Therefore, possible colour reconnections across separate nucleon–nucleon interactions are not considered. Other more recent developments in string models, such as colour ropes and string shoving, are also not considered in \pya. While both effects would be interesting to study in the near future, the goal here is rather to demonstrate the strong sensitivity of flattenicity to the collision impact parameter. 

\begin{figure*}[ht]
\centering
\includegraphics[width=0.45\textwidth]{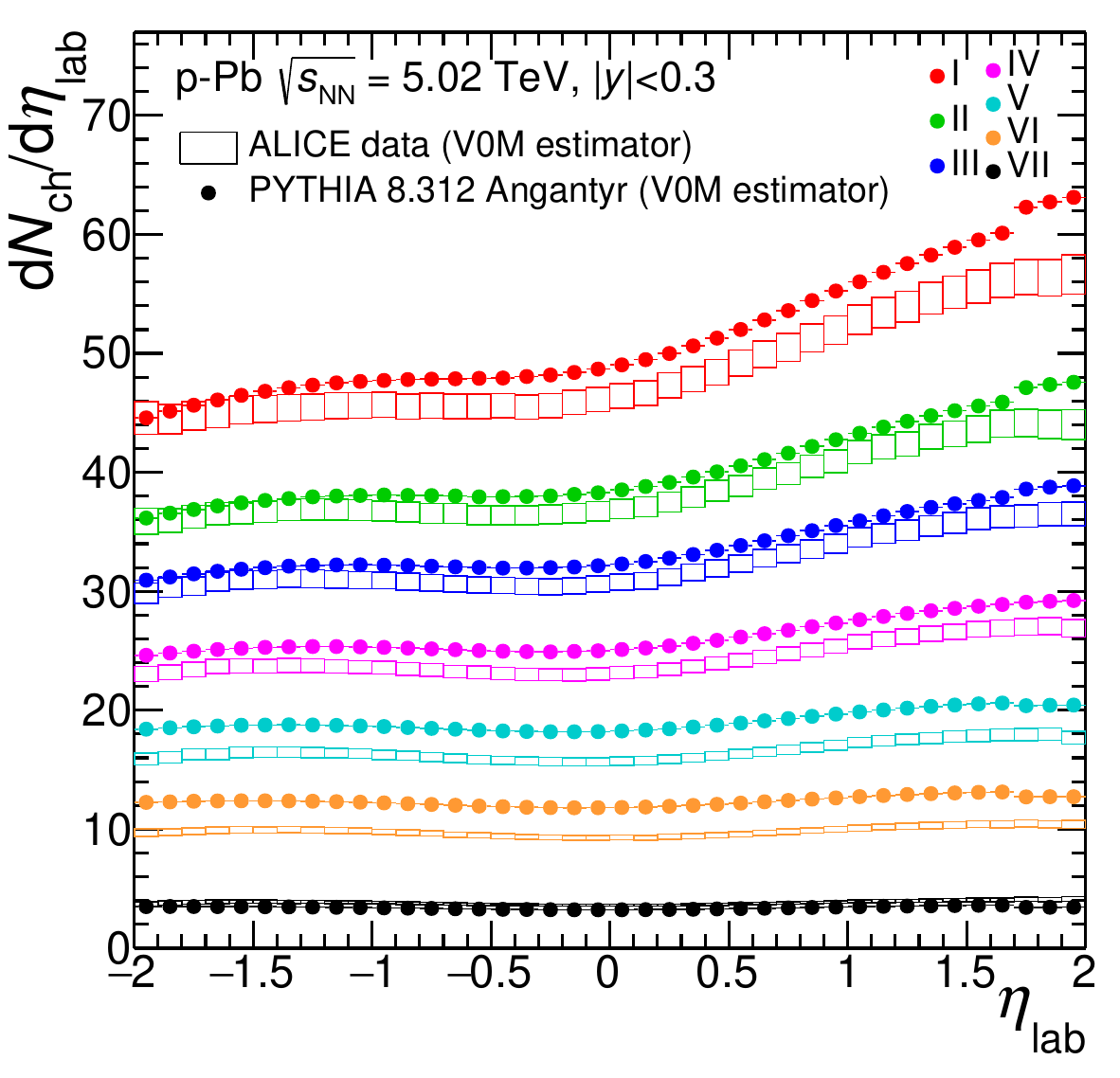}
\includegraphics[width=0.45\textwidth]{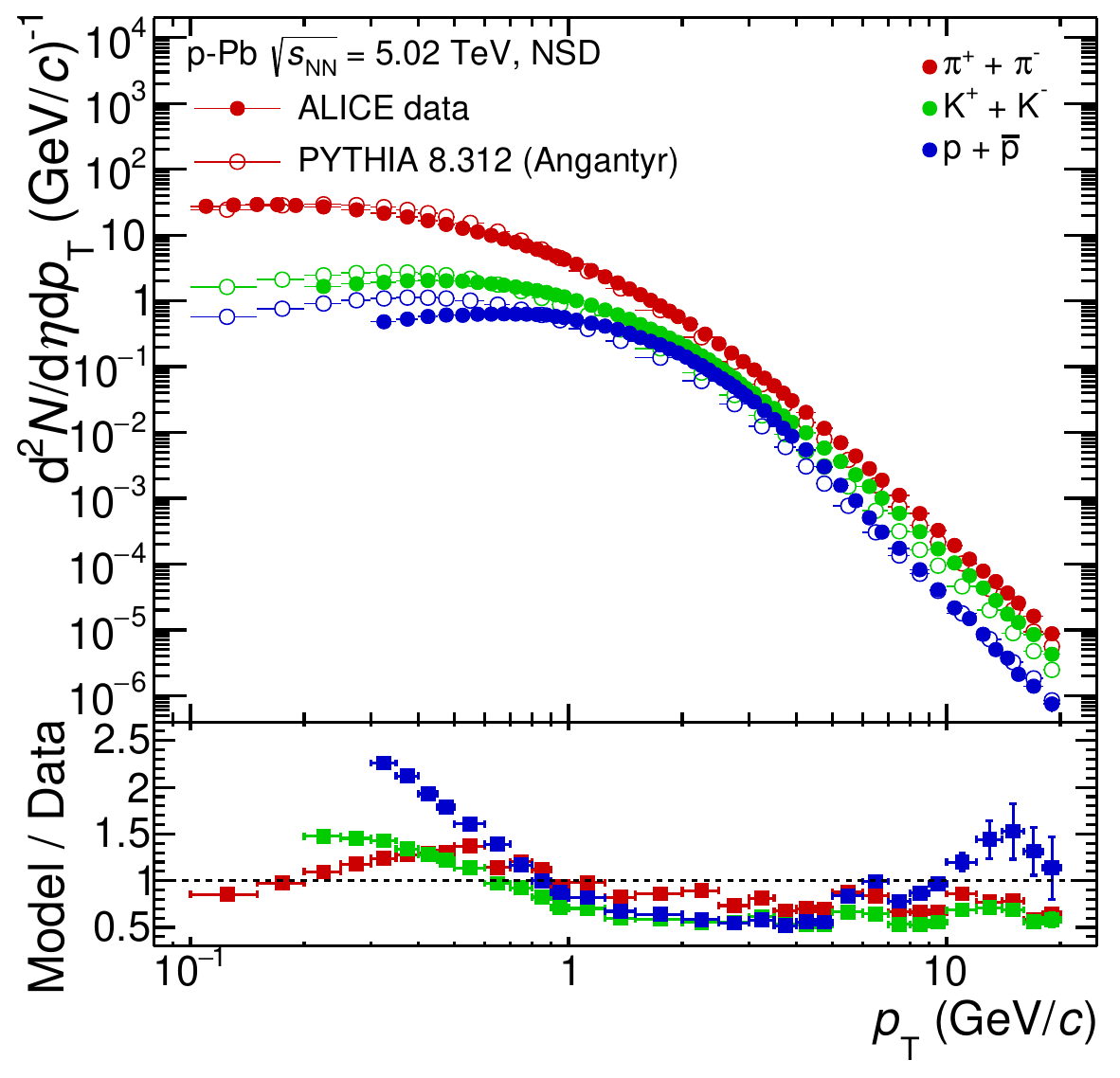}
\caption{Charged-particle density as a function of pseudorapidity ($\eta_{\rm lab}$) (left panel) and transverse momentum spectra of charged pion, kaon and (anti)protons (right panel) for NSD p--Pb collisions at $\sqrt{s}_{\rm NN}=5.02$\,TeV simulated with \pya for different multiplicity classes. Simulations are compared with data measured by the ALICE Collaboration~\cite{ALICE:2016dei}.}
\label{fig:1}  
\end{figure*}

In order to validate the simulations, comparisons with existing data are presented~\cite{ALICE:2014xsp,ALICE:2016dei}.
The left panel of Fig.~\ref{fig:1} shows the average charged-particle multiplicity as a function of pseudorapidity measured for multiplicity classes defined in terms of the forward multiplicity (V0M). Results from the ALICE collaboration are compared with the model predictions. The \pya model describes the data within 20\% for all the multiplicity classes. A similar agreement between ATLAS data and \py was reported in Ref.~\cite{Bierlich:2018xfw}. The right panel of Fig.~\ref{fig:1} shows the transverse momentum distributions of identified light-flavoured hadrons for non-single diffractive (NSD) p--Pb collisions. Although the spectral shapes are qualitative described over eight orders of magnitude, the ratio data-to-model shows that the model has to be significantly improved to reproduce the \pt-differential particle yields. In particular, the model underestimates the particle yields by about 50\% for $p_{\rm T}>2$\,GeV.  

\section{\label{sec:4}Results and discussion}

The ``centrality'' classification using quantities at hadron level (V0M, CL1 and $1-\rho$) are compared with that using the impact parameter of the collision as given by \pya. Seven ``centrality'' classes are defined as indicated in Table~\ref{tab:3}. The lowest-multiplicity (``peripheral'' collisions) class is labelled as VII, whereas the highest-multiplicity (``central'' collisions) class is labelled as I.

\begin{table*}
\caption{\label{tab:3}Multiplicity classes based on different event activity estimators, their corresponding midrapidity charged-pion density ($|y|<0.3$) is presented.}
\begin{tabular}{ l|c|c|c|c|c|c|c }
\hline
\textbf{Mult. class} & \textbf{I} & \bf{II} & \bf{III} & \bf{IV} & \bf{V} & \bf{VI} & \bf{VII} \\
\hline
\textbf{Percentile} & \textbf{0-5\%} & \bf{5-10\%} & \bf{10-20\%} & \bf{20-40\%} & \bf{40-60\%} & \bf{60-80\%} & \bf{80-100\%} \\
\hline
& \multicolumn{7}{c}{$b$ selection} \\
\hline
$\langle{\rm d}N_{\pi} /{\rm d}y\rangle$ &  29.2 & 28.4 & 27.0 & 24.9 & 20.3 & 14.0 & 5.98  \\
$\langle N_{\rm coll}\rangle$ &  12.3 & 11.9 & 11.3 & 9.77 & 7.20 & 4.39 & 1.98   \\
\hline
& \multicolumn{7}{c}{$1-\rho$ selection} \\
\hline
$\langle{\rm d}N_{\pi}/ {\rm d}y\rangle$ &  40.2 & 33.9 & 29.7 & 23.0 & 19.3 & 13.3 & 4.28  \\
$\langle N_{\rm coll}\rangle$ &  13.0 & 12.0 & 10.9 & 9.23 & 6.99 & 4.71 & 1.93   \\
\hline
& \multicolumn{7}{c}{V0M selection} \\
\hline
$\langle{\rm d}N_{\pi} /{\rm d}y\rangle$ &  46.0 & 36.3 & 30.5 & 23.8 & 17.4 & 11.4 & 3.19  \\ 
$\langle N_{\rm coll}\rangle$ &  22.0 & 14.7 & 11.5 & 8.34 & 5.78 & 3.85 & 1.68  \\
\hline
& \multicolumn{7}{c}{CL1 selection} \\
\hline
$\langle{\rm d}N_{\pi} /{\rm d}y\rangle$ &  45.9 & 34.9 & 28.5 & 20.8 & 14.0 & 8.29 & 3.07  \\
$\langle N_{\rm coll}\rangle$ &  16.9 & 12.3 & 10.1 & 7.70 & 5.60 & 3.67 & 1.91   \\
\hline
\end{tabular}
\end{table*}

\begin{figure}[t]
\includegraphics[width=0.49\textwidth]{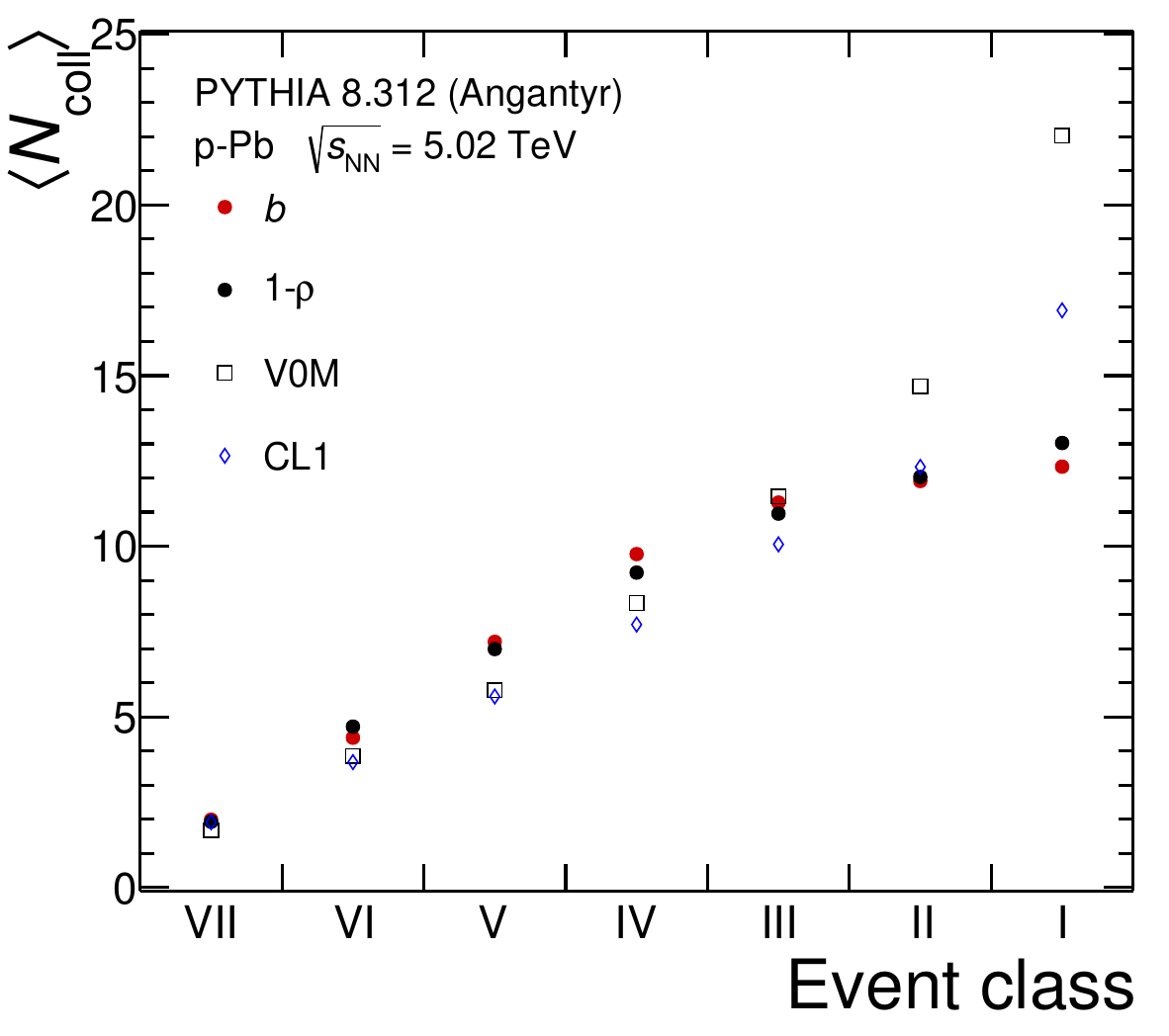}
\caption{Average number of binary nucleon-nucleon for the different multiplicity classes in p--Pb collisions at $\sqrt{s}_{\rm NN}=5.02$\,TeV. The event classification is done using three event activity estimators: flattenicity ($1-\rho$), forward multiplicity (V0M), and midrapidity multiplicity (CL1). Note that in p--Pb collisions $N_{\rm coll}=N_{\rm part}-1$.}
\label{fig:2}  
\end{figure}

Figure~\ref{fig:2} shows the average number of binary nucleon-nucleon collisions for the different ``centrality'' classes. The percentile bins were obtained for four cases: generated impact parameter ($b$), flattenicity ($1-\rho$), forward multiplicity (V0M), and midpidity charged particle density (CL1). For the percentile based on $b$, the dependence is rather weak for the I-III ``centrality'' classes. The average numbers of binary nucleon-nucleon collisions stay within 11-12, for the classes I-III. The variation of around 10\% is similar to that reported by ALICE, although, the $\langle N_{\rm coll} \rangle$ values are larger than the ones reported here. On the other hand, a strong ``centrality'' dependence is observed going from the ``centrality'' class VII to IV. The ``centrality'' dependence of $\langle N_{\rm coll} \rangle$ is quantitatively recovered for the selection based on flattenicity. For multiplicity-based estimators (V0M and CL1) the ``centrality'' dependence is the largest one. In particular for the highest multiplicity class, the deviation is about a factor 2. This observation suggests that centrality based on multiplicity is very sensitive to fluctuations in the number of participant (wounded) nucleons, and therefore it does not correlate very well with the actual impact parameter. However, flattenicity significantly mitigates the effects.

\begin{figure*}[ht]
\centering
\includegraphics[width=0.95\textwidth]{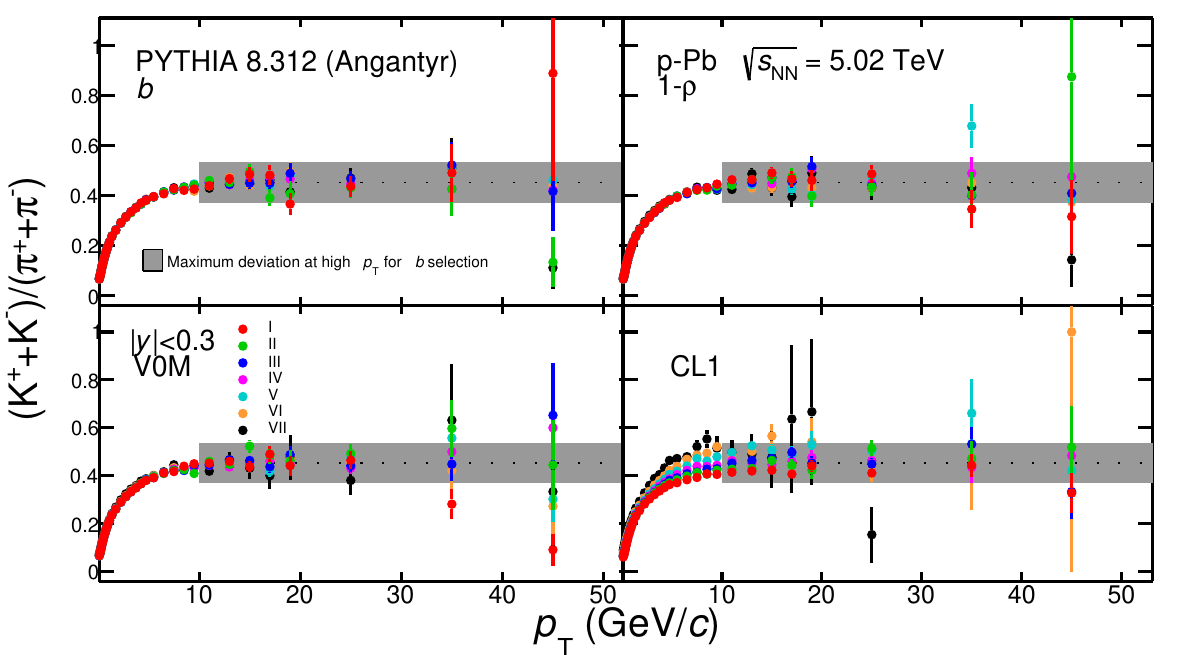}
\includegraphics[width=0.95\textwidth]{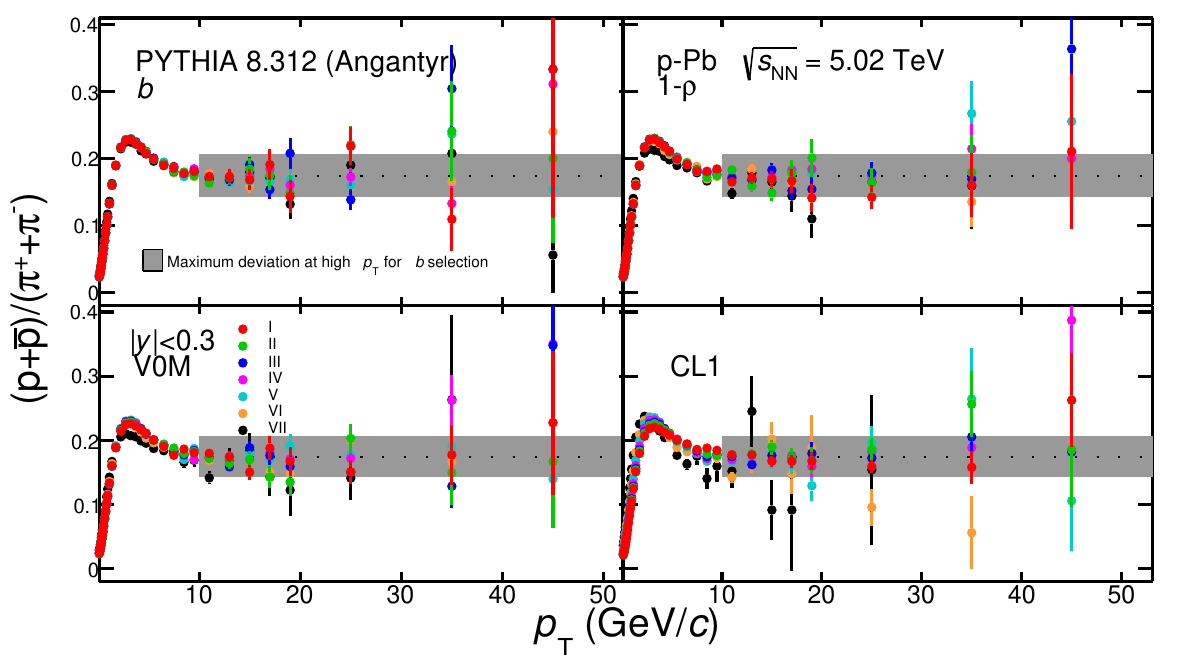}
\caption{Event activity dependence of particle ratios as a function of $p_{\rm T}$. The kaon-to-pion and the proton-to-pion ratios are shown in the upper and bottom panels, respectively. The event classification is done using three event activity estimators: flattenicity ($1-\rho$), forward multiplicity (V0M), and midrapidity multiplicity (CL1).}
\label{fig:3}  
\end{figure*}

Particle ratios as a function of transverse momentum are sensitive to collective-like effects and new hadronization mechanisms like recombination. For example, in pp collisions simulated with PYTHIA, colour reconnection produces flow-like effects in pp collisions with large number of multiparton interactions~\cite{OrtizVelasquez:2013ofg}. The proton-to-pion and kaon-to-pion ratios have been measured as a function of V0M multiplicity in pp, p--Pb and Pb--Pb collisions. For the three colliding systems, the proton-to-pion ratio exhibits a depletion at low transverse momentum ($p_{\rm T}<2$\,GeV/$c$) and an enhancement at intermediate \pt (2-8\,GeV/$c$), both with increasing the event activity. The kaon-to-pion ratio is nearly event activity independent~\cite{ALICE:2018pal}. The studies have been further extended using transverse spherocity~\cite{Ortiz:2015ttf,ALICE:2023bga}. For high-event activity events, the proton-to-pion ratio exhibits a bump (depletion) for isotropic (jet-like) pp collisions  at intermediate \pt. The kaon-to-pion ratio also gets suppressed for jet-like events. Figure~\ref{fig:3} shows the particle ratios for the different ``centrality'' estimators in p--Pb collisions. The ratios are reported for the different ``centrality'' classes. The kaon-to-pion ratio are ``centrality'' independent for the event classification based on impact parameter, flattenicity and forward multiplicity. Whereas, the ratio decreases going from the highest to the lowest CL1 multiplicity class. This behaviour is similar to that reported by ALICE in pp collisions as a function of transverse spherocity~\cite{ALICE:2023bga}, which is in agreement with the fact that requiring high CL1 multiplicity biases the sample towards hard pp collisions. 

Regarding the proton-to-pion ratio as a function of \pt, for all ``centrality'' estimators the flow-like peak is seen at intermediate \pt. This feature is probably a consequence of the colour reconnection mechanisms that is implemented in the nucleon-nucleon interactions. None of the ``centrality'' estimators produce a strong event-activity dependence at intermediate \pt probably because of the lack of colour reconnecting partons from distant nucleon-nucleon interactions, which is a mechanism that is currently under development by \pya authors. It is worth mentioning that CL1 gives a ``centrality'' dependence which yields a suppression of the peak going from peripheral to central p--Pb collisions. This effects is probably a consequence of the presence of jets at midrapidity as observed in the spherocity analysis reported by ALICE~\cite{ALICE:2023bga}.

\begin{figure*}[ht]
\centering
\includegraphics[width=0.95\textwidth]{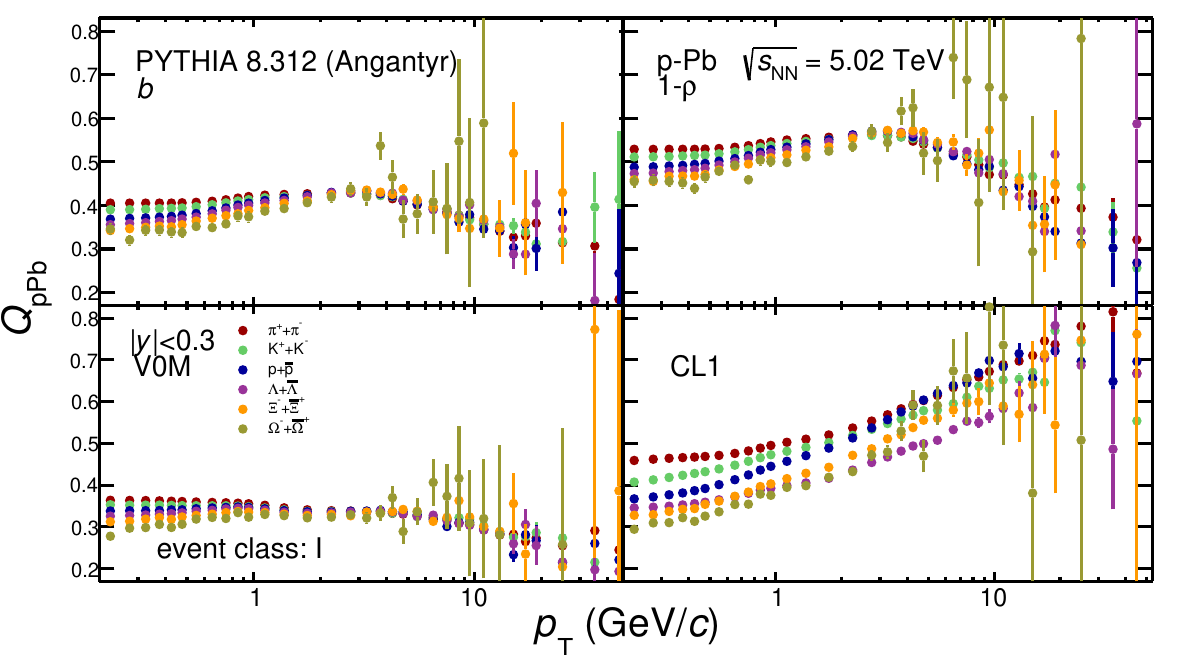}
\caption{Although \pya has issues to reproduce the \pt spectral shape in p--Pb collisions, and therefore, it is known to have difficulties to get the high-\pt yield scaling with $N_{\rm coll}$. The figure shows the particle species dependence of $Q_{\rm pPb}$ as a function of $p_{\rm T}$ for the event class I. The event classification is done using three event activity estimators: flattenicity ($1-\rho$), forward multiplicity (V0M), and midrapidity multiplicity (CL1).}
\label{fig:4}  
\end{figure*}

Since \pya is not able to reproduce the \pt spectral shapes in p--Pb collisions, one would therefore expect that the model will fail at reproducing the scaling of the high-\pt particle yield with the number of binary nucleon-nucleon collisions. However, it is still interesting to see how the different ``centrality'' estimators bias the \pt spectral shapes. To this end, the $Q_{\rm pPb}$ factor is defined as follows~\cite{ALICE:2014xsp}:

\begin{equation}
    Q_{\rm pPb}(p_{\rm T})=\frac{{\rm d}N^{\rm pPb}/{\rm d}p_{\rm T}}{\langle N_{\rm coll} \rangle {\rm d}N^{\rm pp}/{\rm d}p_{\rm T}},
\end{equation}

where $N_{\rm coll}$ is directly obtained from \pya. As discussed in this Ref.~\cite{ALICE:2014xsp}, $Q$ is used instead of $R$ to remember the reader that centrality selection in small systems is strongly affected by biases given the weak correlation between impact parameter and the event activity.

The particle species dependent analysis for the most ``central'' p--Pb collisions is shown in Fig.~\ref{fig:4}. The $\langle N_{\rm coll} \rangle$ values are those provided by the generators, which are listed in table~\ref{tab:3}. From the figure it is clear that the scaling to $N_{\rm coll}$ has to be worked out. However, beyond the normalisation that is a known issue, the discussion can be focused to the particle species dependence of $Q_{\rm pPb}$. The studies reported here include the identified particle yields ($\pi^{+}+\pi^{-}$, ${\rm K}^{+}+{\rm K}^{-}$, ${\rm p}+{\rm \bar{p}}$, $\Lambda+\bar{\Lambda}$, $\Xi^{-}+\bar{\Xi}^{+}$, and $\Omega^{-}+\bar{\Omega}^{+}$) calculated at midrapidity ($|y|<0.3$). For the event selection based on both impact parameter and flattenicty, the \Qppb exhibits a ``bump'' structure in the \pt interval $2<p_{\rm T}<8$\,GeV/$c$, the position of the peak is slightly shifted to higher \pt values with the hadron mass increase. For higher \pt the \Qppb quantity seems to converge to a constant value for all the particle species presented in this paper. The mass effect is more visible at lower \pt ($p_{\rm T}<2$\,GeV/$c$, there, the \Qppb exhibits a depletion going from the lowest multiplicity class (VII) to the highest multiplicity class (I). Qualitatively, this behaviour has been observed in the identified hadron $R_{\rm pPb}$ measured in NSD p--Pb collisions ALICE data~\cite{ALICE:2016dei}. There, for $p_{\rm T}>10$\,GeV/$c$, all nuclear modification factors are consistent with unity within uncertainties. At around $p_{\rm T}\approx4$\,GeV/$c$, the unidentified charged hadron $R_{\rm pPb}$ is above unity, the $R_{\rm pPb}$ for (anti)proton is about 3 times larger than that for charged particles, while for charged pions and kaons the $R_{\rm pPb}$ is below that of charged particles. 

The features of \Qppb discussed above diminishes when V0M or CL1 are used as ``centrality'' estimators. In particular, no bump structure is observed at intermediate \pt. Moreover, the CL1 yields \Qppb that increases with increasing \pt. This is a consequence of the bias towards hard physics that become relevant in small systems.

\section{\label{sec:5}Conclusions}

Flattenicity is a multiplicity estimator that has been implemented in pp data at the LHC energies. Since it considers information both in the azimuthal angle and pseudorapidity, by construction it helps to reduce the biases from local multiplicity fluctuations. Based on \py simulations, flattenicity is expected to be strongly correlated with multiparton interactions, and therefore, with the impact parameter of the collision. Given the sensitivity of flattenicity to the impact parameter, this paper explores its sensitivity to impact parameter in p--Pb collisions. The studies were conducted using the \pya model, which does not assumed a hot thermal QCD system. The goal of the study is to test the ``centrality'' selection based on flattenicity, and compared it with that achieved using other ``centrality'' estimators reported by the ALICE collaboration (V0M and CL1).  The identified particle production for the different estimators is presented with the aim to explore the effects of multiparton interactions and colour reconnection, which the model implements at nucleon-nucleon interaction level. The main observations are listed below.   
\begin{itemize}
    \item The comparison with ALICE data, shows that the charged-particle density as a function of pseudorapidity for different centrality classes based on the V0M estimator is well reproduced by \pya within 20\%. However, the transverse momentum distributions of identified particles in NSD p--Pb collisions exhibit a discrepancy (data to model), which might reach 50\% for $p_{\rm T}>2$\,GeV/$c$.
    \item The average number of binary nucleon-nucleon collisions as a function of the collision ``centrality'' is calculated defining percentiles in impact parameter. The variation on $N_{\rm coll}$ is very weak for the 0-20\% centrality classes. For more peripheral p--Pb collisions, a strong centrality dependence is found. Interestingly, the results obtained using flattenicity as ``centrality'' estimator quantitatively agree with those obtained in terms of impact parameter. The ``centrality'' estimators based on V0M and CL1 exhibit a significant deviation with respect to the results obtained using flattenicity and impact parameter.
    \item Since MPI and CR are both implemented at nucleon-nucleon interaction level meaning that possible reconnections among separate nucleon–nucleon interactions are  neglected, the CR effects originated form individual nucleon-nucleon interactions are studied using the proton-to-pion ratio as a function of \pt. The ratio exhibits a bump structure at intermediate \pt that does not change with collision centrality. This is understood given that the current CR model in \pya does not include a dependence on the p--Pb impact parameter.
    \item Last but not least, the \Qppb factor as a function of \pt was studied. Given the difficulties of \pya to reproduce the \pt spectra in p--Pb collisions, the \Qppb at high \pt does not scale with $N_{\rm coll}$. However, the \Qppb exhibits a strong mass dependence at intermediate \pt when the ``centrality'' selection is done in terms of impact parameter or flattenicity. This behaviour qualitatively agrees with that reported by the ALICE collaboration. The ``centrality'' estimators based on V0M and CL1 are affected from biases due to local multiplicity fluctuations, and in this case, the effects diminishes.
\end{itemize}

Future studies using LHC data as well as comparisons with updated versions of \pya incorporating overlapping strings (colour ``ropes'') will be crucial to make a progress on the understanding of the QGP-like effects observed in small-collisions systems.

\begin{acknowledgments}
Support for this work has been received from CONAHCyT under the Grant CF No. 2042; PAPIIT-UNAM under the project IG100524; as well as from Hungarian National Research, Development and Innovation Office under the Grants OTKA PD143226, FK131979, K135515, NKFIH NEMZ KI-2022-00031 and Wigner Scientific Computing Laboratory (WSCLAB, the former Wigner GPU Laboratory). Authors acknowledge Christian Bierlich for providing the settings to simulate p--Pb collisions in Angantyr and for his useful comments. Authors also acknowledge the technical support of Luciano D\'iaz and Eduardo Murrieta for the assistance to run simulations at the ICN-UNAM computing farm.
\end{acknowledgments}

\bibliography{flat}

\end{document}